\documentclass[aps,prl,superscriptaddress,twocolumn,
groupedaddress,
nofootinbib,
nobalancelastpage,nobibnotes]{revtex4}

\usepackage{amsmath}
\usepackage{amssymb}
\usepackage{amsfonts}
\usepackage{fleqn}

\def\<{\left\langle}
\def\>{\right\rangle}
\def\SnI{N}

\newcommand{\bi}{\begin{itemize}}
\newcommand{\ei}{\end{itemize}}

\newcommand{\be}{\begin{equation}}
\newcommand{\ee}{\end{equation}}
\newcommand{\bea}{\begin{eqnarray}}
\newcommand{\eea}{\end{eqnarray}}

\def\<{\left\langle}
\def\>{\right\rangle}

\newcommand{\SuperField}[1]{\hat{#1}}

\begin{document}

\title{Sneutrino Hybrid Inflation in Supergravity}

\author{Stefan Antusch}
\affiliation{School of Physics and Astonomy, 
University of Southampton, Southampton, SO17 1BJ, United Kingdom} 

\author{Mar Bastero-Gil}
\affiliation{Departamento de Fisica Teorica y del Cosmos 
and Centro Andaluz de Fisica de Particulas Elementales (CAFPE),\\
Universidad de Granada, E-19071 Granada, Spain}

\author{Steve F.~King}
\affiliation{School of Physics and Astonomy, 
University of Southampton, Southampton, SO17 1BJ, United Kingdom}

\author{Qaisar Shafi}
\affiliation{Bartol Research Institute, University of Delaware, 
Newark, DE 19716, USA}

\vspace*{0.2cm}

\begin{abstract}
\vspace*{0.2cm} We propose a hybrid inflation scenario in which the
singlet sneutrino, the superpartner of the right-handed neutrino, plays the
role of the inflaton.  We study a minimal model of sneutrino
hybrid inflation in supergravity, 
where we find a spectral index $n_\mathrm{s}
\approx 1 + 2 \gamma$ with $|\gamma |\lesssim 0.02$, 
and predict a running spectral index 
$|\mathrm{d} n_\mathrm{s}/ \mathrm{d} \ln k| \ll |\gamma| $ 
and a tensor-to-scalar ratio $r \ll \gamma^2$
for field values well below the Planck scale. In our
scenario, the baryon asymmetry of our universe can be explained via
non-thermal leptogenesis and a low reheat temperature $T_\mathrm{RH}
\approx 10^{6}$ GeV can be realized.

\end{abstract}


\maketitle

{\em Introduction} --- Although inflation 
has become a widely accepted paradigm for solving the flatness 
and horizon problems of the
early universe, the relation of inflation to particle physics
remains unclear. Even though there are many models of inflation, 
there are not so many particle physics candidates for the
scalar field responsible for inflation, the so called inflaton
field \cite{Guth:1980zm}. 
The main reason for this is that the inflaton is required
to satisfy slow-roll conditions which are in general difficult
to reconcile with the known couplings of particle physics candidates.
The experimental discovery of neutrino mass and mixing, 
when combined with the ideas of 
the see-saw mechanism \cite{King:2003jb}
and supersymmetry \cite{Chung:2003fi}, gives a new
perspective on this problem. 
In order to generate the observed neutrino masses within 
a see-saw extended version of the minimal supersymmetric standard model
(MSSM), right-handed neutrinos are typically introduced and small neutrino
masses arise naturally from the see-saw mechanism.  Among the
particles of this extended MSSM, the singlet sneutrinos, 
the superpartners of the 
right-handed neutrinos, become 
attractive candidates, in fact the only candidates, for playing the
role of the inflaton.
Motivated by such considerations, the possibility of chaotic (large
field) inflation with a sneutrino inflaton \cite{Murayama:1992ua} has
recently been revisited \cite{Ellis:2003sq}. 
An alternative to chaotic inflation is hybrid inflation \cite{Linde:1993cn}, 
which, in contrast to chaotic inflation, involves field 
values well below the Planck scale.
Hybrid inflation is thereby promising for connecting inflation to particle 
 physics \cite{Dvali:1994ms}.

In this Letter we suggest that one (or more) of the 
singlet sneutrinos $\tilde{\SnI}_i$, where $i=1,2,3$
is a family index, could be the inflaton of hybrid inflation. 
We present a minimal model of sneutrino
hybrid inflation in supergravity, and find the 
spectral index $n_\mathrm{s} \approx 1 + 2 \gamma$ 
with $|\gamma| \lesssim 0.02$, 
$|\mathrm{d} n_\mathrm{s}/ \mathrm{d} \ln k| \ll |\gamma|$ 
and the tensor-to-scalar ratio
 $r \ll \gamma^2$. 
We shall furthermore show how the baryon asymmetry of our universe can be 
explained via non-thermal leptogenesis \cite{Fukugita:1986hr,nonthermalLG}. 
A low reheat temperature 
$T_\mathrm{RH} \approx 10^6$ GeV, 
consistent with the gravitino constraints in some supergravity 
theories \cite{gravitinoproblem}, 
can be realized with values of the neutrino Yukawa couplings 
consistent with first family 
quark and lepton Yukawa couplings in Grand Unified Theories (GUTs).
 
\vspace{2mm}
{\em The Model} ---
Consider the superpotential \cite{footnote2}
 \begin{eqnarray}\label{eq:W4}
\!\!\!\!\!\mathcal{W} = 
\kappa \SuperField{S} \left(\frac{\SuperField{\phi}^4}{{M'}^2} -
M^2\right) \!+
\frac{(\lambda_{N})_{ij}}{M_*}
\,\SuperField{\SnI}_{i}\SuperField{\SnI}_{j}\,\SuperField{\phi}\SuperField{\phi} \, 
+ \dots \,,
\end{eqnarray}
where $\kappa$ and $(\lambda_{N})_{ij}$ are dimensionless Yukawa
couplings and $M,{M'}$ and $M_*$ are three independent mass scales. 
The superfields $\SuperField{\SnI}_i$, $\SuperField{\phi}$ and 
$\SuperField{S}$ contain the 
following bosonic components, respectively: 
the singlet sneutrino inflaton $\tilde{\SnI}$ \cite{footnote},
which is non-zero during inflation;  
the so-called waterfall 
field $\phi$, which is held at zero during inflation
but which develops a non-zero 
vacuum expectation value (vev) after inflation; 
and the singlet field $S$ which is held at zero during and after inflation
\cite{footnote2b}.
The form of $\mathcal{W}$ in Eq.~(\ref{eq:W4}) can be understood
as follows.
The first term on the right-hand side serves to fix 
the vev of the waterfall field after inflation and contributes a large
vacuum energy to the potential during inflation.  
We have chosen the
waterfall superfield to appear in this term as
$\SuperField{\phi}^4/{M'}^2$ instead of 
$\SuperField{\phi}^2$ in order to allow 
a $Z_4$ discrete symmetry to prevent explicit singlet (s)neutrino
masses \cite{domainwalls}. $\mathcal{W}$
is also compatible with a $U(1)_{\mathrm{R}}$-symmetry under
which $\mathcal{W}$ and $\SuperField{S}$ each carry unit R-charge,
while the charge of $\SuperField{\SnI}$ is $1/2$. Under suitable
conditions the discrete subgroup of this symmetry acts as matter
parity \cite{Dvali:1997uq}. 
The second term on the right-hand side of Eq.~(\ref{eq:W4})
allows the sneutrino inflaton to give a positive mass 
squared for the waterfall field during inflation, 
which fixes its vev at zero as long as 
$|\tilde{\SnI}|$ is above a critical value. After inflation, when the 
waterfall field acquires its non-zero vev, the same term
yields the masses of the singlet (s)neutrinos. 

With non-zero F-terms during inflation, the K{\"a}hler potential can 
contribute significantly to the scalar potential. 
Since the field values of the inflaton are well below the reduced 
Planck scale $m_\mathrm{P} =1/\sqrt{8\pi G_\mathrm{N}}$, 
we can consider an expansion in powers of $1/m_\mathrm{P}^2$
\cite{footnote3}:
\begin{eqnarray}\label{eq:K1}
\!\!\!\!\!\!\!\!\!\!\!\!
\lefteqn{ \mathcal{K}  = 
|\SuperField{S}|^2 \!+ |\SuperField{\phi}|^2 \!+ |\SuperField{\SnI}|^2 
\!+\kappa_S \frac{|\SuperField{S}|^4}{4 m_\mathrm{P}^2} 
\!+\kappa_N \frac{|\SuperField{\SnI}|^4}{4 m_\mathrm{P}^2} 
\!+\kappa_\phi \frac{|\SuperField{\phi}|^4}{4 m_\mathrm{P}^2} 
}
\nonumber \\
\!\!\!\!\!\!\!\!\!\!\!\!\!\!\!\!
& \, +& \!\!\kappa_{S\phi} \frac{|\SuperField{S}|^2 |\SuperField{\phi}|^2}{m^2_\mathrm{P}}
\!+ \kappa_{S N} \frac{|\SuperField{S}|^2 |\SuperField{\SnI}|^2}{m^2_\mathrm{P}}
\!+\kappa_{N \phi} \frac{|\SuperField{\SnI}|^2 |\SuperField{\phi}|^2}{m^2_\mathrm{P}}
\!+ \dots \, ,
\end{eqnarray}
where the dots indicate higher order terms and
additional terms for the other fields in the theory. 
With a non-canonical K{\"a}hler potential 
as above, the field $S$ acquires a large mass which holds it 
at zero during inflation.  
We will neglect radiative
corrections to the potential in the following, which are generically  
subdominant in our model. 

\vspace{2mm}
{\em The Potential} ---
We now analyze the scalar potential for the model defined by the 
superpotential $\mathcal{W}$ of Eq.\ (\ref{eq:W4}) and the K{\"a}hler 
potential $\mathcal{K}$ 
of Eq.\ (\ref{eq:K1}).
The F-term contributions to  
the scalar potential are given by: 
\begin{eqnarray}
V_\mathrm{F} = e^{\mathcal{K}/m^2_\mathrm{P}} \left[ 
K_{ij}^{-1} D_{z_i} \mathcal{W} D_{z^*_j} \mathcal{W}^* 
- 3 m_\mathrm{P}^{-2} |\mathcal{W}|^2 
\right] , 
\end{eqnarray}
with $z_i$ being the bosonic components of the superfields
$\SuperField{z}_i \in 
\{\SuperField{\SnI},\SuperField{\phi},\SuperField{S},\dots\}$   
\cite{potential} and where we have defined   
\begin{eqnarray}
D_{z_i} \mathcal{W} := \frac{\partial  \mathcal{W}}{\partial z_i} + 
 m_\mathrm{P}^{-2} \frac{\partial  \mathcal{K}}{\partial z_i} \mathcal{W} 
 \: , \; K_{ij} := \frac{\partial^2 \mathcal{K}}{\partial z_i \partial z^*_j} 
\end{eqnarray}
and $D_{z^*_j} \mathcal{W}^* := (D_{z_j} \mathcal{W})^*$.
Since we assume that 
 $\tilde{\SnI}, \phi$
 and $S$ are effective gauge singlets at the energy scales under consideration, 
 there are no relevant D-term contributions.   
 From Eqs.\ (\ref{eq:W4}) and (\ref{eq:K1}), 
 with canonically normalized fields, 
 we obtain the potential 
\begin{eqnarray}\label{eq:scalarpot1}
V &=& 
\kappa^2 \left(\frac{|\phi|^4}{{M'}^2} - M^2 \right)^2 
\left(1 +(1-\kappa_{S\phi}) \frac{|\phi|^2}{m_\mathrm{P}^2} \right.
+\nonumber \\
&&+\left.\;(1-\kappa_{S N}) \frac{|\tilde{\SnI}|^2}{m_\mathrm{P}^2} 
- \kappa_S \frac{ |S|^2 }{m_\mathrm{P}^2}\right)
\nonumber \\ 
&&+ \; \frac{4 \lambda_N^2}{M_*^2} \,(|\tilde{\SnI}|^4 |\phi|^2 + 
|\tilde{\SnI}|^2 |\phi|^4) +\dots \; ,
\end{eqnarray}
where we have 
shown only the leading order terms and the terms essential for our analysis.

\vspace{2mm}
{\em Sneutrino Hybrid Inflation} ---
Writing the potential of Eq.~(\ref{eq:scalarpot1}) in terms of real fields 
$\tilde{\SnI}_\mathrm{R}=\sqrt{2}|\tilde{\SnI}|$, 
$\phi_\mathrm{R}= \sqrt{2} |\phi|$ and $S_\mathrm{R}=\sqrt{2}|S|$, 
we obtain  
\begin{eqnarray}\label{eq:scalarpot2}
V &=& 
\kappa^2 \left(\frac{\phi_\mathrm{R}^4}{4 {M'}^2} - M^2 \right)^2 
\left(
1 -\beta \frac{ \phi_\mathrm{R}^2}{2 m_\mathrm{P}^2} 
+\gamma \frac{\tilde{\SnI}_\mathrm{R}^2}{2 m_\mathrm{P}^2} 
\right. \nonumber \\
&&\!\!\!\!\!\!\!\!\!- \left. \kappa_S \frac{ S_\mathrm{R}^2 }{2 m_\mathrm{P}^2}  \right) + \frac{\lambda_N^2}{2 M_*^2} \,(\tilde{\SnI}_\mathrm{R}^4 \phi_\mathrm{R}^2 
+ \tilde{\SnI}_\mathrm{R}^2 \phi_\mathrm{R}^4) + \dots  
\,, 
\end{eqnarray} 
where we have defined 
\begin{eqnarray}
\label{eq:beta} \beta &:=&\kappa_{S\phi}-1 \quad \mbox{($>0$ for inflation to
end)}\; ,\\
\gamma &:=& 1-\kappa_{S N}\;.
\end{eqnarray}
During inflation, the waterfall field $ \phi_\mathrm{R}$ has a zero vev and the
potential is dominated by the vacuum energy $V_0 = \kappa^2 M^4$. 
This false vacuum during inflation is stable as long as the mass squared for the waterfall
field $\phi_\mathrm{R}$ is positive. From Eq.~(\ref{eq:scalarpot2}) we obtain
 the requirement  
\begin{eqnarray}
m_{\phi_\mathrm{R}}^2 = 
\lambda_N^2 \frac{\tilde{\SnI}_\mathrm{R}^4}{M_*^2} - 
\beta \frac{\kappa^2 M^4 }{m_\mathrm{P}^2} \;>\; 0\; .
\end{eqnarray}
Inflation ends when $m_{\phi_\mathrm{R}}^2$ becomes negative, i.e.\ 
$\phi_\mathrm{R}$ develops a
tachyonic instability and rolls rapidly to its global minimum at 
$\<\phi_\mathrm{R}\> = \sqrt{2 {M'} M}$. 
Clearly, this requires $\beta > 0$, as already indicated in
Eq.~(\ref{eq:beta}). 
More precisely, inflation ends by a second
order phase transition when the field value of the inflaton drops 
below the critical value $\tilde{\SnI}_{\mathrm{R} c}$ given by 
\be \label{eq:nucrit}
\tilde{\SnI}_{\mathrm{R} c}^2=  \sqrt{\beta}\frac{\kappa}{\lambda_N}\frac{
M^2  M_*}{m_\mathrm{P}} \;.  
\label{phinuc}
\ee
From Eq.~(\ref{eq:scalarpot2}) we see that $S_\mathrm{R}$ can be set to zero 
during inflation 
if we take e.g.\  $\kappa_S < -1/3$, such that $S_\mathrm{R}$ gets a mass term 
larger than the Hubble parameter $H \approx \sqrt{V_0}/(\sqrt{3}m_\mathrm{P})$. 
With $\phi_\mathrm{R} = S_\mathrm{R} = 0$, the part of
the scalar potential relevant for the evolution of the singlet sneutrino 
inflaton $\tilde{\SnI}_\mathrm{R}$ during inflation is given by
\begin{eqnarray}\label{eq:scalarpot3}
V &=& \kappa^2 M^4 \left(
1 
+\gamma \frac{\tilde{\SnI}_\mathrm{R}^2}{2 m_\mathrm{P}^2} 
+ \delta \frac{\tilde{\SnI}_\mathrm{R}^4}{4 m_\mathrm{P}^4}\right) + \dots \;  , 
\end{eqnarray}
where we have included the next-to-leading order term proportional to 
$\delta =
\tfrac{1}{2} + \kappa_{S N}^2 - \kappa_{S N} \kappa_{N} + \tfrac{5}{4} \kappa_{N}$ + \dots \ . 
The parameter $\gamma$ in the scalar potential controls the mass of the 
inflaton. 
Furthermore, compared to the term proportional to $\gamma$, the term proportional
to $\delta$ is suppressed by $\tilde{\SnI}_\mathrm{R}^2/m_\mathrm{P}^2$ 
and will be neglected. 
The slow-roll parameters are 
given by  
\begin{eqnarray}
\!\!\!\!\!\! \epsilon \!&:=&\!  
 \frac{m_\mathrm{P}^2}{2} \!\left(\frac{V'}{V}\right)^2
\!\!\!\simeq \frac{(\delta \tilde{\SnI}_\mathrm{R}^3 \!+ m_\mathrm{P}^2 \gamma  
\tilde{\SnI}_\mathrm{R})^2}{2 m_\mathrm{P}^6}
\approx \gamma^2 \!\frac{\tilde{\SnI}_\mathrm{R}^2}{2 m_\mathrm{P}^2} , \\
\!\!\!\!\!\!\eta \!  &:= &\!   m_\mathrm{P}^2\!\left(\frac{V''}{V}\right) 
\simeq \gamma + \frac{3 \,\delta \tilde{\SnI}_\mathrm{R}^2}{m_\mathrm{P}^2}
\approx \gamma \, , \\
\!\!\!\!\!\!\xi \!  &:=&\!   m_\mathrm{P}^4 \!\left(\frac{V'\, V'''}{V^2}\right)
 \simeq \frac{6\, \delta \tilde{\SnI}_\mathrm{R}^2 (\gamma m_ \mathrm{P}^2 + 
 \delta \tilde{\SnI}_\mathrm{R}^2)}{m_\mathrm{P}^4} ,
\end{eqnarray}
where prime denotes derivative with respect to $N_R$.  
Thus, assuming that the slow-roll approximation  
is justified (i.e.\ $\epsilon \ll 1$, $\eta \ll 1$), the spectral index 
$n_\mathrm{s}$, the tensor-to-scalar ratio
$r=A_\mathrm{t}/A_\mathrm{s}$ and the running spectral index 
$\mathrm{d} n_\mathrm{s}/\mathrm{d} \ln k$ are given by 
\begin{eqnarray}
n_\mathrm{s} \!&\simeq&\! 1 - 6 \epsilon + 2 \eta \;\approx\; 1 + 2 \gamma\; ,
\vphantom{\frac{1}{1}}\\
r \!&\simeq&\! 16 \epsilon \approx  \gamma^2 \, 
\frac{8 \,\tilde{\SnI}_{\mathrm{R}e}^2}{ m_\mathrm{P}^2}\; , \\
\frac{\mathrm{d} n_\mathrm{s}}{\mathrm{d} \ln k} \!\!&\simeq&\!\! 
 16 \epsilon \eta - 24 \epsilon^2 - 2 \xi 
\approx 
-\gamma \, \frac{12  \delta\,\tilde{\SnI}_{\mathrm{R}e}^2}{ m_\mathrm{P}^2}
\; .
\end{eqnarray}
In the above formulae, $\tilde{\SnI}_{\mathrm{R}e}$ is the field value of 
the inflaton $\tilde{\SnI}_\mathrm{R}$ at 
$N=$ 50 to 70 $e$-folds before the end of inflation, given approximately by  
\begin{equation}\label{eq:Nuat60Ne}
\tilde{\SnI}_{\mathrm{R}e} \;\approx\; \tilde{\SnI}_{\mathrm{R} c} 
\,e^{\gamma N }\,,
\label{phinue}
\end{equation}
with the critical value $\tilde{\SnI}_{\mathrm{R} c}$ at the end of inflation defined in Eq.\ (\ref{eq:nucrit}).
The experimental data on the spectral index from WMAP $n_\mathrm{s}=0.99 \pm
0.04$ \cite{Spergel:2003cb} restricts 
$\gamma$ to be roughly $|\gamma| \lesssim 0.02$. As discussed above, $\gamma$
controls the sneutrino mass during inflation. 
In this model it stems mainly from
supergravity corrections. Realizing a very small value of $\gamma$ would 
require some tuning. 
In addition, we see that
the tensor-to-scalar ratio
$r=A_\mathrm{t}/A_\mathrm{s}$ and the running spectral index 
$\mathrm{d} n_\mathrm{s}/\mathrm{d} \ln k$ are suppressed by higher powers of
$\gamma$ or by $\tilde{\SnI}_\mathrm{R}^2/m_\mathrm{P}^2$ and are thus generically small.
Especially the prediction for the tensor-to-scalar ratio $r \ll \gamma^2$ 
is thus in sharp contrast to the prediction of $r=0.16$ for the case of 
chaotic sneutrino inflation.

In our model, the amplitude of the 
primordial spectrum is given by 
\begin{equation}
P_{\cal R}^{1/2} \simeq 
\frac{1}{\sqrt{2 \varepsilon}} \left(\frac{H}{2\pi
m_{\mathrm{P}}}\right)
\approx \frac{\kappa}{2 \sqrt{3} \,\gamma \,\pi}
\frac{M^2}{m_\mathrm{P}\, \tilde{\SnI}_{\mathrm{R}e}} \,.
\label{spectrum}
\end{equation} 
 Given the COBE normalization $P_{\cal R}^{1/2}\approx 5 \times 10^{-5}$ 
 \cite{Smoot:1992td},
from Eqs.\ (\ref{phinuc}), 
 (\ref{spectrum}) and (\ref{phinue}) we obtain 
\begin{equation}\label{eq:scales}
\frac{M^2}{M_*\,m_\mathrm{P}} \approx \:3 \times 10^{-8} \,
\frac{\gamma^2\,\sqrt{\beta}}{\kappa \, \lambda_N}\;,
\end{equation}
which relates the scale $M$ in the superpotential to the cutoff scale $M_*$. 
It has to be combined with the constraint 
$\tilde{\SnI}_{\mathrm{R}e}\ll m_\mathrm{P}$
 (see Eqs.~(\ref{eq:Nuat60Ne}) and (\ref{eq:nucrit})) and with 
 $M < {M'}, M_*$.  

\vspace{2mm}
{\em Reheating and Non-thermal Leptogenesis} --- 
In our scenario, the observed baryon asymmetry can arise 
via non-thermal leptogenesis \cite{Fukugita:1986hr,nonthermalLG}. 

Let us assume the situation that the inflaton is the lightest
singlet sneutrino $\tilde{\SnI}_1$ and that it dominates leptogenesis and reheating 
after inflation \cite{warning}. This is e.g.\ the case if the waterfall field 
$\phi$  
decays earlier than the singlet sneutrino inflaton via heavier 
singlet neutrinos $\SnI_2$ (or $\SnI_3$) with comparably 
large couplings to $\phi$. From
Eq.~(\ref{eq:W4}), using $\<\phi \> = \sqrt{{M'} M}$, we see
that its mass is given by $M_\mathrm{R1} = 2 (\lambda_N)_{11} {M'}M/M_*$ in
the basis where the mass matrix $M_\mathrm{R}$ of the singlet (s)neutrinos is 
diagonal. 
It decays mainly via the extended MSSM
Yukawa coupling $(Y_\nu)_{i1}
\SuperField{L}_{i} \SuperField{H}_\mathrm{u} \SuperField{\SnI}_1$ into
slepton and Higgs or into lepton and Higgsino with a decay width given
by $\Gamma_{\SnI_1} = M_\mathrm{R1} {(Y^\dagger_\nu Y_\nu)_{11}}/{(4\pi)}$. 
The decay of the singlet sneutrino after inflation reheats the universe to a
temperature $T_{\mathrm{RH}} \approx (90/(228.75\pi^2 ))^{1/4} \sqrt{
\Gamma_{\SnI_1} m^{}_{\mathrm{P}}}$. 
If $M_\mathrm{R1}\gg T_{\mathrm{RH}}$, the lepton asymmetry is produced
via cold decays of the singlet sneutrinos \cite{Campbell:1992hd}. In this case,
the produced baryon asymmetry can be estimated as
${n_\mathrm{B}}/{n_\gamma} \approx - 1.84 \, \varepsilon\,
{T_{\mathrm{RH}}}/{M_\mathrm{R1}}$, where $\varepsilon$ is the decay
asymmetry for the singlet sneutrino decay. 
Note that $\varepsilon$ is bounded
by $|\varepsilon| \lesssim \tfrac{3}{8 \pi} \, \sqrt{\Delta m^2_{31}}
{M_\mathrm{R1}}/{v^2_\mathrm{u}}$
\cite{Hamaguchi:2001gw}  
for the case of hierarchical
singlet (s)neutrinos and light
neutrinos. $\Delta m^2_{31} \approx 2.6 \times 10^{-3}$ eV$^2$ is
the atmospheric neutrino mass squared difference and 
$v_\mathrm{u} = \<H_\mathrm{u}\>$.  
The bound on $\varepsilon$ implies
$|{n_\mathrm{B}}/{n_\gamma}| \lesssim 
1.84 \, \sqrt{\Delta m^2_{31}} T_{\mathrm{RH}} \,3/{(8 \pi v^2_\mathrm{u})}$, and hence
$T_{\mathrm{RH}} \gtrsim 10^6$ GeV for the observed baryon-to-photon
ratio $n_\mathrm{B} /n_\gamma = (6.5^{+0.4}_{-0.8}) \times 10^{-10}$
\cite{Spergel:2003cb}. 

To take a concrete example of the above discussion, neutrino Yukawa
couplings $(Y_\nu)_{i1}\approx 10^{-6}$ and a sneutrino mass of
$M_\mathrm{R1}=10^8$ GeV imply a reheat temperature
$T_{\mathrm{RH}} \approx 10^6$ GeV, compatible with the gravitino
constraints (see e.g.\ \cite{gravitinoproblem}) in some supergravity
theories, saturating the lower bound on $T_{\mathrm{RH}}$
\cite{footnote4}. In this example
the lightest singlet neutrino is effectively decoupled from
the see-saw mechanism as in sequential dominance \cite{Antusch:2004gf}. 
Using $M_\mathrm{R1} = 2 (\lambda_N)_{11} {M'}M/M_*$
and Eq.~(\ref{eq:scales}), taking $\gamma = \beta = 10^{-2}$
yields $\kappa = 10^{-2} {M'}/M$.
In addition,
$\tilde{\SnI}_{\mathrm{R}e}\ll m_\mathrm{P}$ is satisfied for $M
{M'}/m^2_\mathrm{P} \ll 10^{-4}$
(using Eqs.~(\ref{eq:Nuat60Ne}) and (\ref{eq:nucrit})).
We see that this can be achieved
easily with $M \approx 0.1 {M'}$ and both scales
somewhat below the GUT scale. $M_*$ is not constrained directly and can for example
be around $10^{17}$ GeV. With $M \approx 10^{15}$ GeV and
$M' \approx 10^{16}$ GeV we obtain 
for instance $\tilde{\SnI}_{\mathrm{R}e}\approx 10^{16}$ GeV, 
well below the Planck Scale. For 
$(\lambda_N)_{33}={\cal O}(1)$, 
the heaviest singlet (s)neutrino has a mass 
$M_{\mathrm{R}3} \approx 10^{14}$ GeV in this case.

\vspace{2mm}
{\em Summary and Conclusions} --- 
We have proposed a hybrid inflation scenario
in which one (or more) of the singlet sneutrinos, the superpartners of the 
right-handed neutrinos, play the role of the inflaton. 
Sneutrinos are present in any extension of the MSSM where the smallness of the
observed neutrino masses is explained via the see-saw mechanism. 
In a minimal model of sneutrino
hybrid inflation in supergravity
we have shown how the baryon asymmetry of our universe can be 
explained via non-thermal leptogenesis after inflation. 
For achieving a low reheat temperature, 
the Yukawa couplings can have values consistent with first family 
quark and lepton Yukawa couplings in GUT models. For example,  
the inflaton can be the lightest singlet sneutrino with a mass around $10^8$
GeV and can have Yukawa couplings 
$\approx 10^{-6}$ in order to realize a reheat temperature 
$T_\mathrm{RH} \approx 10^6$ GeV, 
consistent with the gravitino constraints in some supergravity theories. 
In contrast to
chaotic inflation, the field values of the singlet sneutrino inflaton in hybrid
inflation are well
below the Planck scale, so that the supergravity corrections can be carefully
monitored. 
In the minimal model considered here these corrections
play an essential role. 
We have found the 
spectral index $n_\mathrm{s} \approx 1 + 2 \gamma$ with 
$|\gamma |\lesssim 0.02$
and a running spectral index $|\mathrm{d} n_\mathrm{s}/
\mathrm{d} \ln k| \ll |\gamma|$.
 Furthermore, sneutrino hybrid inflation predicts  
 a small tensor-to-scalar ratio $r \ll \gamma^2$, 
 much smaller than the prediction $r \approx 0.16$ of chaotic sneutrino
 inflation. This makes sneutrino hybrid inflation   
 easily distinguishable from chaotic sneutrino inflation.

\section*{Acknowledgements}  
We would like to thank Nefer Senoguz and Mark Hindmarsh
for very helpful discussions.  
We acknowledge support from the PPARC grant PPA/G/O/2002/00468 and from 
DOE under contract DE-FG02-91ER40626.

\providecommand{\bysame}{\leavevmode\hbox to3em{\hrulefill}\thinspace}


\begin{thebibliography}{10}
\bibitem{footnote}
Henceforth we will consider one of the singlet sneutrinos as the
inflaton for simplicity and drop family indices.
\bibitem{domainwalls}
The dots in Eq.~(\ref{eq:W4}) include $Z_4$-violating    
higher-dimensional operators such as, e.g., 
$\SuperField{S} \SuperField{\phi}^5/M'^3$ 
(or even $\SuperField{S} \SuperField{\phi}^5/m_\mathrm{P}^3$)  
which lift the degeneracy 
of the true vacuum and effectively blow away potential domain wall networks 
associated with $Z_4$-breaking after inflation.
\end{thebibliography}
\end{document}